# Nanocrystals in silicon photonic crystal standing-wave cavities as spin-photon phase gates for quantum information processing

Y.-F. Xiao

*Optical Nanostructures Laboratory, Columbia University, New York, NY 10027 and Key Laboratory of Quantum Information, University of Science & Technology of China, Hefei 230026, P. R. China*

J. Gao, X. Yang, and R. Bose

*Optical Nanostructures Laboratory, Columbia University, New York, NY 10027*

Guang-Can Guo

*Key Laboratory of Quantum Information, University of Science & Technology of China, Hefei 230026, P. R. China*

C. W. Wong[*]

*Optical Nanostructures Laboratory, Columbia University, New York, NY 10027*

By virtue of a silicon high-$Q$ photonic crystal nanocavity, we propose and examine theoretically interactions between a stationary electron spin qubit of a semiconductor nanocrystal and a flying photon qubit. Firstly, we introduce, derive and demonstrate the explicit conditions towards realization of a spin-photon two-qubit phase gate, and propose these interactions as a *generalized* quantum interface for quantum information processing. Secondly, we examine single-spin-induced reflections as direct evidence of intrinsic bare and dressed modes in our coupled nanocrystal-cavity system. The excellent physical integration of this silicon system provides tremendous potential for large-scale quantum information processing.

[*] Corresponding author: cww2104@columbia.edu

Cavity quantum electrodynamics (QED) provides an almost ideal system for implementing quantum information and computation [1-5]. Recently, photonic crystal nanocavities with high quality factors ($Q$) and ultrasmall mode volumes are attracting attention in the context of optical cavity QED [6-9]. Combined with low-loss and strong localization, they present a unique platform for highly integrated nanophotonic circuits on a silicon chip, which can also be regarded as a quantum hardware for nanocavity-QED-based quantum computing. Toward this goal, strong interactions between a quantum dot and a single photonic crystal cavity have been observed experimentally [6-9]. Moreover, single photons from a quantum dot coupled to a source cavity can be remarkably transferred to a target cavity via an integrated waveguide in an InAs/GaAs solid-state system [10], which opens the door to construct the basic building blocks for future chip-based quantum information processing systems. Weak coupling nanocrystal ensemble measurements were reported in $TiO_2$-$SiO_2$ and AlGaAs cavity systems (below 1 μm wavelengths) recently [11-12] and also independently in silicon nanocavities with lead chalcogenide nanocrystals at near 1.55 μm fiber communication wavelengths recently [13]. In this Letter, we examine the single-photon pulse (or weak coherent light pulse) interactions of a single semiconductor nanocrystal in a system comprised of *standing-wave* high-$Q/V$ silicon photonic crystal nanocavities. In contrast to earlier traveling-wave whispering gallery mode cavity studies [14], we show here that a QED system based on coupled standing-wave nanocavities can realize a spin-photon phase gate even under the bad-cavity limit and provide a *generalized* quantum interface for quantum information processing. In addition, we demonstrate numerically a solid-state universal two-qubit phase gate operation with a single qubit rotation. This theoretical study is focused within the parameters of near 1.55 μm wavelength operation for direct integration with the fiber network, and in the silicon materials platform to work with the vast and powerful silicon processing infrastructure for large-array chip-based scalability.

We begin by considering a combined system consisting of coupled point-defect high-*Q/V* photonic crystal cavities, a line-defect photonic crystal waveguide, and an isolated single semiconductor nanocrystal. We offer some brief remarks on this system before building our theoretical model. When a photon pulse is coupled into the cavity mode via a waveguide (Fig. 1(a)), photons can couple out of the cavity along both forward and backward propagating directions of the waveguide because the cavity supports standing-wave modes. While each cavity can each have a Faraday isolator to block the backward propagating photon, such implementation may not be easily scalable to a large-array of cavities. To obtain only forward transmission, here we examine theoretically a defect cavity system with accidental degeneracy [15-17] as a generalized study of *cavity-dipole-cavity* systems, and which also provides close to 100% forward-only drop efficiency. This framework is also immediately applicable to non-reciprocal magneto-optic cavities which have larger fabrication tolerances [18]. Both systems support two degenerate even $|e\rangle$ and odd $|o\rangle$ cavity modes (*h*-polarized, dominant in-plane *E*-field) that have opposite parity due to the mirror symmetry, as shown in Fig. 1(a). The waveguides can support both *v*-polarizations (dominant in-plane *H*-field) and *h*-polarizations for polarization diversity [19].

Fig. 1(b) shows the energy levels and electron-exciton transitions of our cavity-dipole-cavity system. In order to produce nondegenerate transitions from the electron spin states, a magnetic field is applied along the waveguide direction [20]. $|\uparrow\rangle$ and $|\downarrow\rangle$ play the rule of a stationary qubit, which have shown much longer coherence time than an exciton (dipole or charge). The transition $|\uparrow\rangle \leftrightarrow |e_1\rangle$, with the descending operator $\sigma_- = |\uparrow\rangle\langle e_1|$, is especially chosen and coupled with the cavity modes with

single-photon coupling strengths $g_e(\vec{r})$ and $g_o(\vec{r})$, while other transitions are decoupled with the cavity modes.

Now we construct our model by studying the interaction between the nanocrystal and the cavity modes. The Heisenberg equations of motion for the internal cavity fields and the nanocrystal are

$$\frac{dc_e}{dt} = -i[c_e, H] - \kappa_e c_e + i\sum_{j=1,2} \sqrt{\kappa_{e1}} c_{in}^{(j)}, \qquad (1)$$

$$\frac{dc_o}{dt} = -i[c_o, H] - \kappa_o c_o + (-1)^{j+1} \sum_{j=1,2} \sqrt{\kappa_{o1}} c_{in}^{(j)}, \qquad (2)$$

$$\frac{d\sigma_-}{dt} = -i[\sigma_-, H] - \gamma \sigma_-, \qquad (3)$$

where the interaction Hamiltonian $H = \delta_{al}\sigma_+\sigma_- + \sum_{p=e,o}[\delta_{cl} c_p^\dagger c_p + g_p(\vec{r}) c_p \sigma_+ + h.c.]$ is in a rotating frame at the input field frequency $\omega_l$. In contrast to earlier work [14, 21-22], here we examine the case with the two $|e\rangle$ and $|o\rangle$ modes in the standing-wave cavities in order for forward-only propagation of the qubit. The cavity dissipation mechanism is accounted for by $\kappa_{e(o)} = \kappa_{e(o)0} + \kappa_{e(o)1}$, where $\kappa_{e(o)0}$ is intrinsic loss and $\kappa_{e(o)1}$ the external loss for the even (odd) mode. The nanocrystal dissipation is represented by $\gamma \equiv \gamma_s/2 + \gamma_p$ where $\gamma_s$ is the spontaneous emission rate and $\gamma_p$ the dephasing rate of the nanocrystal.

When the two degenerate modes have the same decay rate, i.e., $\kappa_{e0} = \kappa_{o0} = \kappa_0$, $\kappa_{e1} = \kappa_{o1} = \kappa_1$, and $\kappa \equiv \kappa_0 + \kappa_1$, two new states $|\pm\rangle = (|e\rangle \pm i|o\rangle)/\sqrt{2}$ are suitable to describe this system, which can be thought as two traveling (or rotating) modes. In this regard, the interaction Hamiltonian is expressed as $H = \delta_{al}\sigma_+\sigma_- + \sum_{s=+,-}[\delta_{cl} c_s^\dagger c_s + g_s(\vec{r}) c_s \sigma_+ + h.c.]$, where the effective single-photon coupling rates are $g_\pm(\vec{r}) = (g_e(\vec{r}) \mp ig_o(\vec{r}))/\sqrt{2}$. In this case, Eqs. (1), (2), and (3) are rewritten into the corresponding forms with $c_\pm$.

The nanocrystal-cavity system is excited by a weak monochromatic field (e.g., single-photon pulse), so that we solve the above motion equations for the below explicit analytical expressions

$$\sigma_-(\omega) = -i \sum_{s=+,-} g_s(\vec{r}) c_s(\omega)/(i\delta_{al} + \gamma) \tag{4}$$

and $c_\pm(\omega)$ are given as

$$i\sqrt{2\kappa_1} c_{in}^{(1)}(\omega) - (i\delta_{cl} + \kappa) c_+(\omega) - ig_+^*(\vec{r}) \sigma_-(\omega) = 0, \tag{5}$$

$$i\sqrt{2\kappa_1} c_{in}^{(2)}(\omega) - (i\delta_{cl} + \kappa) c_-(\omega) - ig_-^*(\vec{r}) \sigma_-(\omega) = 0. \tag{6}$$

Note that orthogonality of the $|e\rangle$ and $|o\rangle$ basis modes (as shown in Fig. 1a) forces the nanocrystal to choose only either $g_e(\vec{r}) = +ig_o(\vec{r})$, or $g_e(\vec{r}) = -ig_o(\vec{r})$, or both (in which case $|e\rangle$ and $|o\rangle$ are uniquely zero), but no other possibilities. Photon qubit input from only the left waveguide forces only one of the cavity states ($|e\rangle + i|o\rangle$) to exist [15], and we assume this cavity environment from the existing photon qubit enhances the $g_e(\vec{r}) = -ig_o(\vec{r})$ probability. Of course, with only the left waveguide qubit input in a non-reciprocal magneto-optic cavity, this condition is strictly enforced. Hence we can take $g_e(\vec{r}) = -ig_o(\vec{r})$, which implies $g_-(\vec{r}) = 0$, $g_+(\vec{r}) = \sqrt{2} g_e(\vec{r})$, to further simplify Eqs. (5)-(6). Now note that the left output $c_{out}^{(1)}$ remarkably vanishes, while the right output is given by $c_{out}^{(2)} = c_{in}^{(1)} (\kappa - 2\kappa_1 - i\delta + \lambda)/(\kappa - i\delta + \lambda)$, where $\lambda = 2|g_e(\vec{r})|^2 /[i(\Delta - \delta) + \gamma]$, and $\Delta \equiv \delta_{al} - \delta_{cl}$ and $\delta \equiv -\delta_{cl}$ denote the nanocrystal-cavity and input-cavity detunings, respectively. Importantly, this implies that our quantum phase gate provides a *true* one-way transmission through the cavity-dipole-cavity system.

To examine more of the underlying physics, we consider first the case of exact resonance ($\Delta = 0, \delta = 0$). When $|g_e(\vec{r})|^2/\kappa\gamma \gg 1$ (the nanocrystal occupies the spin state $|\uparrow\rangle$), we obtain $c_{\text{out}}^{(2)} \approx c_{\text{in}}^{(1)}$. When $g_e(\vec{r}) = 0$ (the nanocrystal occupies the spin state $|\downarrow\rangle$), we obtain $c_{\text{out}}^{(2)} \approx -c_{\text{in}}^{(1)}$ for $\kappa_1 \gg \kappa_o$, which indicates that the system achieves a *global* phase change $e^{i\pi}$. This distinct characteristic allows the implementation of a spin-photon phase gate. After the photon pulse passes though the cavity system, we easily obtain a gate operation

$$\begin{aligned}&|h\rangle|\uparrow\rangle \to |h\rangle|\uparrow\rangle, \ |h\rangle|\downarrow\rangle \to -|h\rangle|\downarrow\rangle, \\ &|v\rangle|\uparrow\rangle \to |v\rangle|\uparrow\rangle, \ |v\rangle|\downarrow\rangle \to |v\rangle|\downarrow\rangle.\end{aligned} \quad (7)$$

This two-qubit phase gate combined with simple single-bit rotation is, in fact, universal for quantum computing. More importantly, this interacting system can be regarded as a quantum interface for quantum state sending, transferring, receiving, swapping, and processing.

To efficiently evaluate the quality of the gate operation, the gate fidelity is numerically calculated, as shown in Fig. 2. Considering specifically a lead chalcogenide (e.g. lead sulphide) nanocrystal and silicon photonic nanocavity system for experimental realization, we choose the spontaneous decay as $\gamma_s \sim 2$ MHz and all non-radiative dephasing $\gamma_p \sim 1$ GHz at cooled temperatures. Photonic crystal cavities have an ultrasmall mode volume $V$ ($\sim 0.1\ \mu m^3$ at 1550 nm), with a resulting calculated single-photon coherent coupling rate $|g_e|$ of $\sim 30$ GHz. High $Q$s of up to even $\sim 10^6$ experimentally and $\sim 10^7$ theoretically [23-24] has been achieved in photonic crystal cavities.

With these parameters, as shown in Fig. 2a, the gate fidelity of the cavity-dipole-cavity system can reach 0.98 or more, even when photon loss is taken into account, and even when the vacuum Rabi frequency $g_e$ is lower than the cavity decay rate $\kappa$ (bad-cavity limit). The gate fidelity increases initially as the cavity approaches more into the over-coupling regime due to less photon loss and eventually decreases as the nanocrystal-cavity system moves away from the strong coupling regime. Secondly, we note that with non-zero detuning ($\Delta/\kappa_o=2$; Case III and VI), the gate fidelity slightly decreases but is still adequate. With increasing nanocrystal dissipation rate (Fig. 2b), the fidelity decreases as expected and the system moves away from strong coupling (less nanocrystal interactions with the cavity). The physical essence behind such high fidelities is the true one-way transmission where the nanocrystal couples to $|+\rangle$ mode, with only forward propagation with no backward scattering of the qubit. In addition, accidental degeneracy mismatch may degrade the gate performance. To validate the feasibility of the present scheme, we perform a direct calculation of gate fidelity for different frequency and lifetime of the opposite-parity cavity modes. Even with degeneracy mismatch ($\omega_e - \omega_l = \delta_{el} \neq \delta_{ol} = \omega_o - \omega_l$; in Case IV and VII) with some backward scattering of the qubit, the gate fidelity is shown to remain high. Moreover, with different lifetimes of the cavity modes (Case VII), the fidelity remains high as long as the $|g(\vec{r})|^2/\kappa\gamma \gg 1$ condition is satisfied.

Furthermore, we show that the above cavity-dipole-cavity interaction mechanism can result in interesting transmissions and reflections based on the presence or absence of dipole interaction, and with different detunings. We examine the case of $g_e(\vec{r}) = g$ and $g_o(\vec{r}) = 0$, such as when the nanocrystal is positioned at the cavity mirror plane. Some typical transmission and reflection spectra are shown in Fig. 3. In the absence of a dipole (i.e., the nanocrystal occupies the spin state $|\downarrow\rangle$), the cavity system has near-unity transmission, except when on-resonance. However, when the located nanocrystal is in state $|\uparrow\rangle$, the interacting system is transmission-free and remarkably reflects the cavity field strongly. We emphasize that this reflection is induced by a *single* spin state, and hence can be termed single-spin-induced reflection. The constructive interference of the cavity field can be considered as an optical-analog to electromagnetically induced absorption in the excited state of a 3-level atomic system [14, 21]. The three reflection peaks in Fig. 3(a) can be understood by considering the strong cavity-dipole-cavity interaction, where the input photon pulse experiences three modes: bare odd mode (central peak) and two dressed even modes (side peaks). When the total cavity decay increases, the three peaks overlap increasingly and form a new peak (Fig. 3b-3d) at zero detuning input. We note the high reflectivity for the cavity-dipole-cavity system at zero detuning, even under the bad-cavity limit. This high reflectivity is helpful to permit arrayed controlled phase flip operation with a single circulator at the input.

In summary, we have theoretically introduced, derived and demonstrated the robust implementation of a single spin-photon phase gate in a cavity-dipole-cavity system, where each localized cavity mode consists of standing-wave photonic crystal point-defect modes. The conditions of accidental degeneracy are examined to enforce complete transfer, either in the forward transmission or in reflection, of the qubit. In addition, we examine the coupled system transmission and reflectivity, where we observe that a photon pulse is strikingly reflected by a cavity interacting with a single spin, even under the bad-cavity limit. This combined nanocrystal-cavity system, implemented in a silicon materials platform with lead chalcogenide nanocrystals in the near-infrared, can serve as a two-qubit phase gate and, indeed, as a general quantum interface for large-array chip-based quantum information processing.

The authors acknowledge funding support from the National Science Foundation (ECS-0622069), DARPA, and the New York State Office of Science, Technology and Academic Research. X. Yang acknowledges support of an Intel Fellowship. Y.-F. Xiao and G.-C. Guo acknowledge funding support from National fundamental Research Program of China.

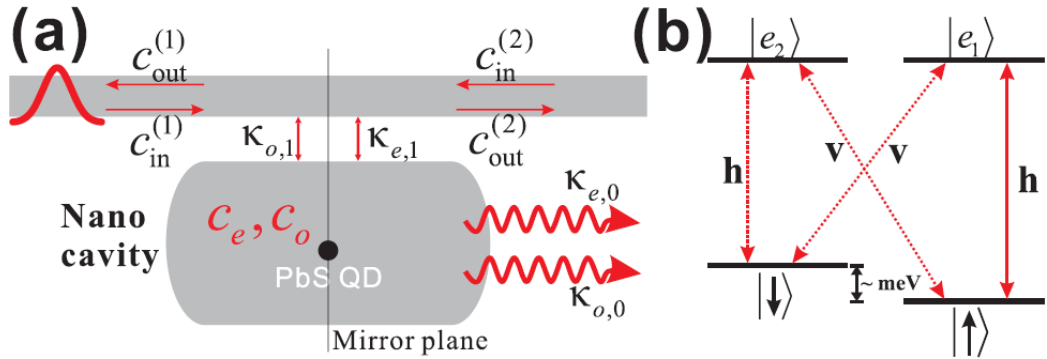

Fig. 1 (Color online) (a) Sketch of a waveguide side coupled to a cavity which supports two degenerate modes $c_e$ and $c_o$ with opposite parity. (b) lead chalcogenide (e.g. lead sulphide) nanocrystal energy levels and the electron-exciton transitions in the presence of a strong magnetic field along the waveguide direction, which produces nondegenerate transitions from the electron spin states $|\uparrow\rangle$ and $|\downarrow\rangle$ to the charged exciton states $|e_1\rangle$ and $|e_2\rangle$ under the transition selection rules.

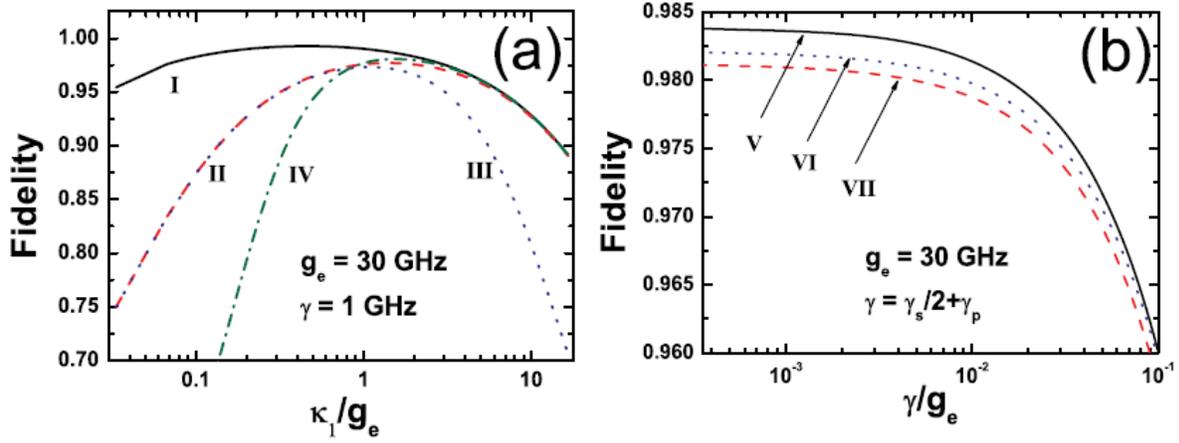

Fig. 2 (Color online) Gate fidelity versus $\kappa_1$ (panel a) and $\gamma$ (panel b) respectively for the lead sulphide nanocrystal in degenerate cavity modes, illustrating that the fidelity mainly depends on $g_e^2/\kappa\gamma$ and $\kappa_0/\kappa_1$. **Case I:** $\kappa_0$ = 0.1GHz, $\delta_{el} = \delta_{ol} = \delta_{al}$ = 0. **Case II:** $\kappa_0$ = 1GHz, $\delta_{el} = \delta_{ol} = \delta_{al}$ = 0. **Case III:** $\kappa_0$ = 1GHz, $\delta_{el} = \delta_{ol}$ = 0, $\delta_{al} = 5\kappa_0$. **Case IV:** $\kappa_{e0} = \kappa_{o0}$ = 0.1GHz, $\delta_{el} = -\delta_{ol}$ = 5GHz, $\delta_{al}$ = 0. **Case V:** $\kappa_{e0} = \kappa_{o0} = \kappa_0$ = 1GHz, $\kappa_{e1} = \kappa_{o1} = \kappa_1 = g_e$, $\delta_{el} = \delta_{ol} = \delta_{al}$ = 0. **Case VI:** identical to Case V but with $\delta_{al} = 5\kappa_0$. **Case VII:** $\kappa_{e0} = 2\kappa_{o0}$ = 0.2GHz, $\kappa_{e1} = 1.1\kappa_{o1} = g_e$, $\delta_{el} = -\delta_{ol}$ = 5GHz, $\delta_{al}$ = 1GHz.

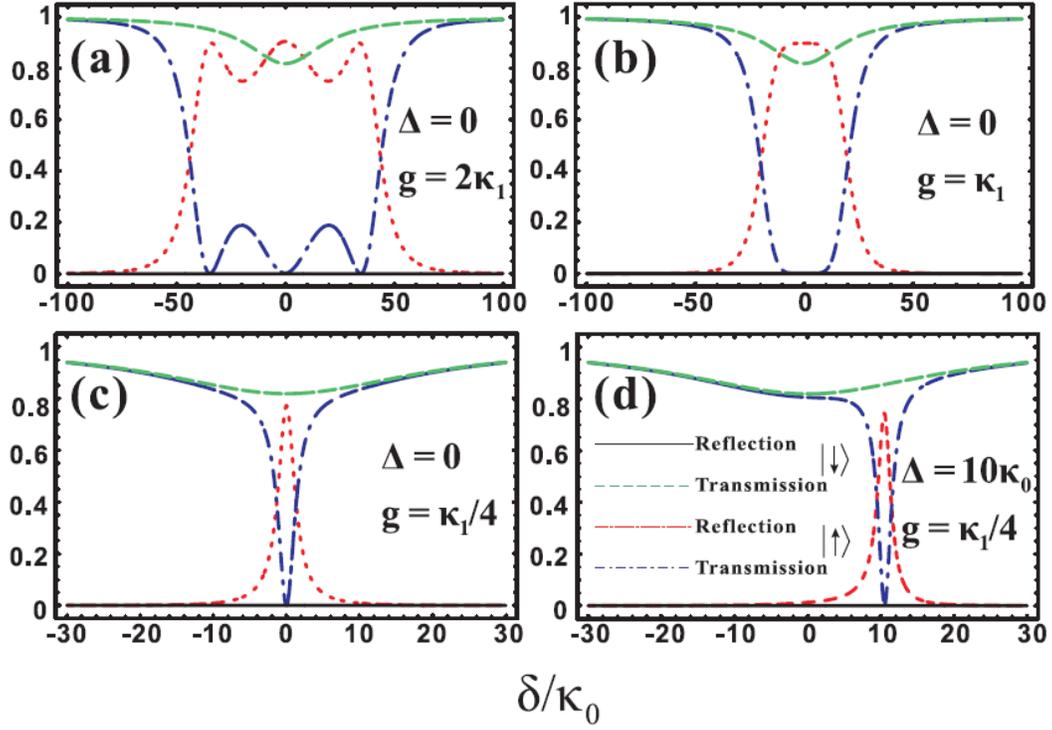

Fig. 3 (Color online) Reflection and transmission of the spin-photon phase gate with an isolated semiconductor nanocrystal in the degenerate point-defect standing-wave cavity modes. Other conditions for this parameter set include: $\gamma = \kappa_0/10$ and $\kappa_1 = 20\kappa_0$, with the nanocrystal located at the cavity mirror plane ($g_e(\vec{r}) = g$ and $g_o(\vec{r}) = 0$). The black solid (green dashed) line is the reflection (transmission) in the absence of a dipole in the cavity. The red dotted (blue dashed-dot) line is the reflection (transmission) in the presence of a dipole in the cavity.